\documentclass[preprint,preprintnumbers,amsmath,amssymb,nofootinbib]{revtex4}

\usepackage{graphicx}
\usepackage{dcolumn}
\usepackage{bm}
\usepackage{color}

\begin{document}
\title{Twist-stretch relations in nucleic acids}

\author{Marco Zoli}

\affiliation{School of Science and Technology \\  University of Camerino, I-62032 Camerino, Italy \\ marco.zoli@unicam.it}

\date{\today}

\begin{abstract}
Nucleic acids are highly deformable helical molecules constantly stretched, twisted and bent in their biological functioning. Single molecule experiments have shown that double stranded (ds)-RNA and standard ds-DNA have opposite twist-stretch patterns and stretching properties when overwound under a constant applied load. The key structural features of the A-form RNA and B-form DNA helices are here incorporated in a three-dimensional mesoscopic Hamiltonian model which accounts for the radial, bending and twisting fluctuations of the base pairs. Using path integral techniques which sum over the ensemble of the base pair fluctuations, I compute the average helical repeat of the molecules as a function of the load. The obtained twist-stretch relations and stretching properties, for short A- and B-helical fragments, are consistent with the opposite behaviors observed in kilo-base long molecules.
\end{abstract}


\maketitle

\section*{1. Introduction}

The mechanical properties of nucleic acids molecules underlie their biological functions as, in vivo, various binding proteins constantly bend, twist, stretch, cut and reseal the helical chains while regulating their activities and gene expression \cite{stasiak97,kalos11,ort14,marko15,biton18,grzy19,lee19}. Single molecule experiments relying on optical and
magnetic tweezers techniques \cite{busta92,block97,mameren09}, have probed the mechanical response of kilo-base long double stranded (ds) DNA sequences to a tunable  load applied along the molecular axis. It has been highlighted that the DNA intrinsic flexibility is governed by entropic elasticity up to forces of $\sim 10$ pN whereas, at larger forces, the molecule is progressively straightened with its end-to-end distance (see Fig.~\ref{fig:1}) becoming of order of the contour length. At $\sim 65$ pN, the molecule is stretched up to $\sim 1.7$ times its contour length due to structural changes of the intra-strand phosphate bonds. \footnotemark{}

\footnotetext{This holds if, in the experimental setup, only two ends of the strands are anchored. Instead, the over-stretching transition is shifted to $\sim 110$ pN when all four ends of the strands are anchored so that the helix is torsionally constrained.
}

This over-stretching transition had been first ascribed to the formation of a distinct double helical S-state \cite{cluzel96,marko99} and then interpreted as a force induced melting of specific domains of the chain \cite{rouz01,maiti09} whereby both processes may also co-exist \cite{nord14,lomb14}. Importantly, it has been shown that kilo-base ds-DNA over-winds as a function of the applied loads up to $\sim 30$ pN  whereas, at larger forces, the helix unwinds. Moreover, the molecule elongates when over-wound under a constant force which is large enough to suppress the bending fluctuations \cite{busta06}. While these findings have fostered research on the peculiar twist-stretch coupling of DNA, several theoretical methods, e.g., molecular dynamics and Monte-Carlo simulations  \cite{tan15}  have suggested that the helix over-twisting/untwisting pattern versus stretching may hold also for chains of a few tens of base pairs. 
More recently, force extensions measurements have become available also for dsRNA \cite{dekker14} thus revealing stretching and twist-stretch properties remarkably different from ds-DNA. Markedly, stretching a RNA duplex causes it to untwist while RNA shortens its helical extension when over-wound under a constant tension. 

As ds-RNA adopts a right-handed  A-type helix, with shorter rise distance and broader diameter than the standard right-handed B-form of ds-DNA, one is led to investigate whether structural helical properties may be at the origin of peculiar twist-stretch patterns which, in turn, affect the nucleic acids interactions with proteins and their biological functioning. Indeed, it has been recognized since long that helical DNA can also exist in the A-form, the transition from the B- to the A-form being driven by the loss of water content in the environment \cite{frank53} \footnotemark{}. 

\footnotetext{As the long term stability of physiological B-DNA is required in a multitude of applications, dehydration conditions are routinely employed e.g, in methods for digital information storage \cite{puddu15}. On the other hand, dehydration may structurally transform DNA, unwind the helix and ultimately lead to unwanted denaturation effects \cite{sena18}.   
}

In this regard, all atom potential energies simulations of DNA helical conformations had previously suggested \cite{olson99} that A- and B- DNA structures may respond differently to external perturbations with A-type and B-type fragments respectively untwisting and over-twisting upon stretching.
Molecular dynamics simulations have shed light also on the mentioned peculiar  properties of dsRNA \cite{oroz10} and it has been argued that such  properties may be ascribed to distinctive structural features, i.e., the base pair stacking and their inclination with respect to the helical axis  \cite{tan16}. 

Here I focus on the interplay between form and helical conformation of nucleic acids \cite{bohr11,gole}, performing a quantitative analysis of their twist-stretch properties based on a
Hamiltonian model which realistically represents the forces stabilizing the double helix and suitably incorporates the mentioned structural features as tunable
microscopic parameters. The use of a mesoscopic Hamiltonian provides a description of the helix at the level of the base pair, a feature which is particularly appropriate at those short length scales in which fluctuational effects are strong and the validity of the traditional worm-like-chain of polymer physics has been questioned \cite{widom,archer06,wigg06,vafa12,maiti15,io16b,lam17}.

Other characteristic quantities measuring the global flexibility of the molecules such as the persistence lengths are not addressed here as they appear substantially similar in the two nucleic acids structures \cite{zachar15} and can be evaluated via models which do not require the application of external loads \cite{io18b,zhang23}.

\section*{2. Model }

\begin{figure}
\includegraphics[height=10.0cm,width=7.5cm,angle=-90]{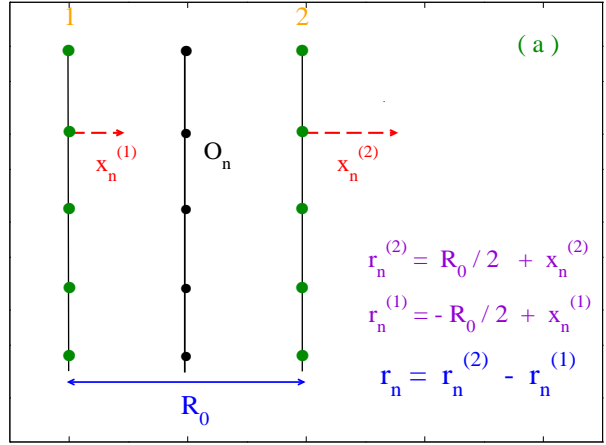}
\includegraphics[height=10.0cm,width=7.0cm,angle=-90]{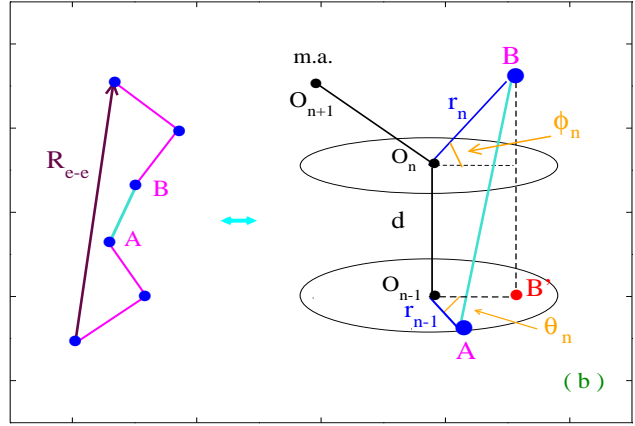}
\caption{\label{fig:1}(Color online)  (a) One dimensional model for an open end chain with $N$ point-like base pairs arranged along the two complementary strands. The transverse base fluctuations, $r_n^{(1,2)}$, are measured with respect to the mid-chain axis hence, the distance $r_n$ between the pair mates is defined with respect to $R_0$ i.e., the bare helix diameter.
(b) Three dimensional model. The global size of the chain is measured by the end-to-end distance, $R_{e-e}$, shown on the l.h.s. drawing \cite{io16a}. The schematic model of one basepair step is shown on the r.h.s. The segment $\overline{AB}$, i.e. the separation between two adjacent fluctuations along the molecular axis (m.a.), is the distance between the tips of the radial displacements $r_{n}$, $r_{n-1}$. 
The $O_n$'s are arranged along the molecular axis at a constant distance $d$.  {$\phi_{n}$ is the bending angle between adjacent $r_{n}$'s in a dimer;  $\theta_n$ is the twist accumulated along the helix whose average  value is calculated via Eq.~(\ref{eq:09}).} In the absence of bending degrees of freedom, the $r_{n}$'s would represent in-plane fluctuations sweeping the ovals in the r.h.s. drawing. 
}
\end{figure}

To begin with, I recall the geometric features of the  model for an open end chain with $N$ point-like base pairs, already employed to calculate the flexibility properties of DNA \cite{io19,io20b}. The schematic of the model is shown in Figs.~\ref{fig:1}. On the {top},  (a) shows the standard one dimensional model \cite{pey89} in which the base pairs (green dots) of equal mass are arranged like beads along two parallel strands set at the distance $R_0$ which represents the bare helix diameter. For each pair mate, only transverse fluctuations are considered and the only degree of freedom is $r_n^{(1,2)}$. Accordingly, 
$r_{n}$ is defined as the inter-strand fluctuation between the complementary  mates of the $n-th$ base pair. Such distance is measured with respect to the point $O_n$ which lies along the central axis of the helix. In the absence of radial fluctuations,  all $r_{n}$'s would be equal to $R_0$  and the model would reduce to a freely jointed chain  made of $N-1$ bonds, all having length $d$.   Fig.~\ref{fig:1}(b) depicts the extension of the model to the three dimensional case which is treated in the following calculations. Here the blue dots represent the tips of the inter-strand fluctuations $r_{n}$'s {while $\theta_n$ is the accumulated twist angle along the helix. It is measured with respect to the twist angle for the preceding base pair in the chain.} In the presence of the twisting angles only, the $r_{n}$'s would represent in-plane fluctuations spanning the ovals in the r.h.s. drawing. In this case, the model would be a two dimensional fixed-plane representation for the helix \cite{barbi99}. Adding the bending fluctuations between adjacent planes, the model becomes three dimensional. Note that $\phi_{n}$ measures the bending fluctuation of the  $n-th$ base pair radial displacement with respect to the preceding one in the dimer.

If the average bending angles are close to zero, then
the $n-th$ base pair radial fluctuations occur in a plane almost perpendicular to the molecular axis. If this feature is common to a large number of dimers in the chain then there is no overall significant tilt of the base pairs planes. This picture is appropriate to model the physiological B-form of ds-DNA, in which the helix axis runs through the center of each base pair and the base pairs are stacked per­pendicular to the axis \cite{io22}. 

Further, $d$ measures the fixed inter-plane distance along the molecular axis while the distance $\overline{d_{n,n-1}}$  between adjacent base pairs, i.e., $\overline{AB}$ in Fig.~\ref{fig:1}(b), is obtained in terms of the fluctuational degrees of freedom. In the absence of fluctuations, $\overline{d_{n,n-1}}$ would reduce to $d$.

\begin{figure}
\includegraphics[height=11.0cm,width=9.0cm,angle=-90]{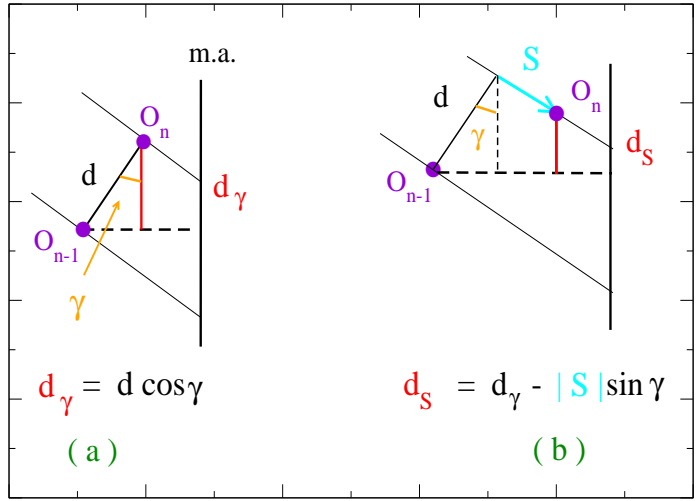}
\caption{\label{fig:2}(Color online)  (a) Base-pair inclination with respect to the helical axis. The base pair planes are not normal to the molecular axis and the  tilt angle $\gamma $ (taken as positive) measures the deviation from the perpendicular configuration. As a result the rise distance (along the molecular axis) $d_{\gamma }$ gets shorter than $d$.
(b) The n-th base pair slides on top of the adjacent (n-1)-th base pair along the stack. In this case, $S$ is taken as negative hence, the rise distance (along the molecular axis) $d_{S }$ gets shorter than $d_{\gamma }$.
}
\end{figure}

The more compact A-form helix adopted by ds-RNA \footnotemark{} 
\footnotetext{ The presence of a hydroxil group bound to the 2' carbon of the ribose ring is the key reason forcing RNA to coil into the A-form \cite{carlo06} 
}
displays two microscopic features, visualized in Fig.~\ref{fig:2}, whose entity is dependent on the di-nucleotide step:  i) the base pair planes are tilted by the angle $\gamma $ respect to the vertical helical axis and ii) the base pairs forming a dimer, slide by a distance $S$ past each other. For simplicity it is hereafter assumed that, for a single simulation, $\gamma $ and $S$ are average values distinctive of the chain although local variations are expected in specific sequences \cite{oroz10}.
The occurrence of tilt and slide has the direct consequence to shorten the rise distance along the helical axis, a feature which affects the helix stretching flexibility \cite{calla}. Accordingly, for the A-form helix,  $\overline{d_{n,n-1}}$ would reduce to $d_S$ in the absence of fluctuations.

Moreover, since adjacent base pairs along the stack are also twisted, due to the slide their center is shifted outwards and displaced to the side of the helical axis \cite{dick83}. Thus, the width of the stack gets larger and the A-form helix has a broader average diameter than the B-form. As the average twist angle is smaller, the A-form helix has a higher helical repeat (i.e., the number of base pairs per helix turn, defined hereafter as $h$ ) than the B-form.

\section*{3. Hamiltonian }

One can assume to stretch the helical molecules, both the B- and A-form, applying a force $F_{ex}$ to one end of the chain while the opposite end is fixed. As the molecule is not torsionally constrained it may change its average helical repeat upon stretching. While different pulling schemes can be devised \cite{romano13}, the model corresponds to an experimental setup in which both helical strands are stretched and the force is tuned in a range of values well below the over-stretching transition \cite{mameren09}.  Also, taking $(F_{ex})_{max} < 30pN$, I consider tensions which cause a kilo-base long DNA fragment to over-twist \cite{busta06}.
As the force opposes the bending fluctuations, the molecular axis is straightened along the pulling direction. It is also assumed that the strength exerted by the load is uniform for all nucleotides arranged along the chain.

Then, the Hamiltonian for a molecule with $N$ base pairs of reduced mass $\mu$, stacked in a helical conformation and stretched by a force $F_{ex}$, is:

\begin{eqnarray}
& &H =\, H_a[r_1] + \sum_{n=2}^{N} H_b[r_n, r_{n-1}, \phi_n, \theta_n] \, , \nonumber
\\
& &H_a[r_1] =\, \frac{\mu}{2} \dot{r}_1^2 + V_{1p}[r_1] \, , \nonumber
\\
& &H_b[r_n, r_{n-1}, \phi_n, \theta_n]= \,  \frac{\mu}{2} \dot{r}_n^2 + V_{1p}[r_n] + V_{2p}[ r_n, r_{n-1}, \phi_n, \theta_n] - F_{ex} d_S  \cos\bigl( \phi_n \bigr)  \, . \nonumber
\\ 
\label{eq:01}
\end{eqnarray}

$d_S$ is defined in Fig.~\ref{fig:2}. In the absence of tilt and slide, $d_S \equiv d$. 

$H_a[r_1]$ is taken out of the sum as the first base pair is coupled only to the successive base pair along the stack. {The angular contributions to the kinetic energy are not considered in Eq.~(\ref{eq:01}) as explained in Section 4.}
$V_{1p}[r_n]$ is one-particle potential modeling the inter-strand forces at the \textit{n-th} site. $V_{2p}[ r_n, r_{n-1}, \phi_n, \theta_n]$ is the two-particle stacking term which accounts for the covalent bonds between adjacent base pairs and incorporates the model dependence on the angular degrees of freedom.
Their explicit expressions are:

\begin{eqnarray}
& &V_{1p}[r_n]=\, V_{M}[r_n] + V_{Sol}[r_n] \, , \nonumber
\\
& &V_{M}[r_n]=\, D_n \bigl[\exp(-b_n (|r_n| - R_0)) - 1 \bigr]^2  \, , \nonumber
\\
& &V_{Sol}[r_n]=\, - D_n f_s \bigl(\tanh((|r_n| - R_0)/ l_s) - 1 \bigr) \, , \nonumber
\\
& &V_{2p}[ r_n, r_{n-1}, \phi_n, \theta_n]=\, K_{n, n-1} \cdot \bigl(1 + G_{n, n-1}\bigr) \cdot \overline{d_{n,n-1}}^2  \, , \nonumber
\\
& &G_{n, n-1}= \, \rho_{n, n-1}\exp\bigl[-\alpha_{n, n-1}(|r_n| + |r_{n-1}| - 2R_0)\bigr]  \, . \nonumber
\\ 
\label{eq:02}
\end{eqnarray}

The Morse potential $V_{M}[r_n]$, depending on the base pair dissociation energy $D_n$ and on the inverse length $b_n$, is usually taken to model hydrogen bonds. It features a hard core accounting for the repulsive interaction between the charged phosphates on complementary strands, a stable minimum and a dissociation plateau. 
Thus, for base pair fluctuations such that,  $|r_n| - R_0 \gg b_n^{-1} $, the pair mates would sample the flat part of the Morse potential and could go infinitely apart with no energy cost. Accordingly $V_{M}[r_n]$ does not describe those strand recombination events which instead take place in solutions, i.e. hydrogen bonds with the surrounding solvent, whose rate depends on the proton concentration. 

{The neglect of the environment was a shortcoming of the one-dimensional ladder model \cite{pey93} affecting mostly the DNA dynamics, e.g. the estimate of the lifetimes of the base pairs open states \cite{pey09}. 
Later on the model has been improved by adding a one-particle solvent potential, $V_{Sol}[r_n]$,   
that introduces a hump  whose width is tuned by $l_s$ and whose maximum sets the energy threshold around which a base pair may first temporarily open and then either
re-close or fully dissociate \cite{coll95,druk01}. Then, the statistical properties of the model can be studied by taking a solvent term which stabilizes the pair interactions enhancing by $f_s D_n$ the height of the energy barrier over the Morse plateau. 
}
{Instead, to account for the environment in the DNA dynamics, other methods directly add a viscous force to the nonlinear equation of motion, derived from the 1D ladder model, for the base pair breathing fluctuations  \cite{zdrav01}. 
}

$V_{2p}$ contains both the elastic force constant $K_{n, n-1}$ for the $n-$th dimer and  the nonlinear parameters $\rho_{n, n-1}$, $\alpha_{n, n-1}$ which account for the cooperativity effects, i.e. formation of local bubbles, observed in the denaturation regime \cite{wart85}. 
While, for $K_{n, n-1}$, the range of appropriate values has been analyzed both experimentally and theoretically  at least in DNA \cite{eijck11,io16b,kalos20}, less information is available as for the nonlinear parameters \cite{io20}.

For $\alpha_{n, n-1}$,  the condition \, $\alpha_{n, n-1}  < b_{n}$ should hold in order to ensure that the range of the stacking potential is broader than that of the Morse potential. This is consistent with the fact that covalent bonds along the stack are stronger than inter-strand base pair hydrogen bonds. 
The physical meaning behind the specific choice for $V_{2p}$ is understood by observing that, when all base pairs are closed, $|r_n| - R_0  \ll \alpha_{n, n-1}^{-1}$ for all $n$. Under these conditions, it is noticed from Eq.~(\ref{eq:02}) that the effective coupling is $ \sim K_{n, n-1}(1 + \rho_{n, n-1})$. 

However, because of a large amplitude fluctuation at the $n-$th site, the inequality $|r_n| - R_0 > \alpha_{n, n-1}^{-1}$ may be fulfilled. When this occurs the
hydrogen bonds between pair mates are disrupted and the base moves out of the stack. Accordingly also the interaction between neighboring bases along the strand weakens, due to a reduced $\pi$ electrons overlap,  and the effective coupling drops to $\sim K_{n, n-1}$. As a consequence, also the adjacent base moves out of the stack thus propagating the fluctuational opening. This explains the relation between non-linearity and cooperativity which causes bubble formation and eventually denaturation of the double helix in the model of Eq.~(\ref{eq:02}).
Then, taking small $\alpha_{n, n-1}$ values,  it is assumed that large fluctuations are required to disrupt a base pair and consistently lower the stacking energy while the
$\rho_{n, n-1}$ parameters weigh the energy difference between a closed and open base pair conformation. 

$V_{2p}$ also incorporates the dependence on the angular degrees of freedom through the base pair distance $\overline{d_{n,n-1}}$ shown in Fig.~\ref{fig:1}(b) \cite{io18c}. In the A-form helix, $\overline{d_{n,n-1}}$ is shortened as the rise distance contracts due to the tilt and slide.

\section*{4. Partition function }

The Hamiltonian in Eqs.~(\ref{eq:01}),~(\ref{eq:02}) can be treated by path integral techniques \cite{fehi} to extract information on the thermodynamical and structural helical properties as described in previous works \cite{io14b}. 
The computational method is based on the idea that the radial fluctuations $r_n$  are trajectories and, accordingly, can be mapped onto the time scale: $r_n \rightarrow |r_n(\tau)|$ \, whereby \, $\tau$ is the imaginary time varying in a range $[0, \beta]$ and $\beta$ is the inverse temperature.  The partition function $Z_N$ is expressed as an integral over closed trajectories, $\,r_n(0)=\, r_n(\beta) \,$,  running along the $\tau$-axis. {Details of the method are found in ref.\cite{kleinert}.}  Importantly, the closure condition for the fluctuations is imposed on the time axis whereas the chain maintains the open ends in real space. This avoids to set boundary conditions which close a chain into a loop as usually done in Transfer Integral methods for simplified one-dimensional DNA Hamiltonian models \cite{zhang97,hando12,albu14,singh15}. Such a procedure however may not be adequate to deal with short open fragments for which boundary effects are relevant.

Then, taking in Eq.~(\ref{eq:01}) the radial fluctuations as paths, $Z_N$ reads:

\begin{eqnarray}
& &Z_N=\, \oint Dr_{1} \exp \bigl[- A_a[r_1] \bigr]   \prod_{n=2}^{N}  \int_{- \phi_{M} }^{\phi_{M} } d \phi_n \int_{- \theta_{M} }^{\theta _{M} } d \theta_{n} \oint Dr_{n}  \exp \bigl[- A_b [r_n, r_{n-1}, \phi_n, \theta_n] \bigr] \, , \nonumber
\\
& &A_a[r_1]= \,  \int_{0}^{\beta} d\tau H_a[r_1(\tau)] \, , \nonumber
\\
& &A_b[r_n, r_{n-1}, \phi_n, \theta_n]= \,  \int_{0}^{\beta} d\tau H_b[r_n(\tau), r_{n-1}(\tau), \phi_n, \theta_n] \, \, ,
\label{eq:32}
\end{eqnarray}

where $\phi_{M}$ and $\theta _{M}$ are the maximum amplitudes for the bending and twisting fluctuations hereafter set to $\pi /15$ and $\pi /2$ respectively \cite{lank10,kim14}.

Consistently with the closure condition, the base pair paths can be expanded in Fourier series, \, $r_n(\tau)=\, (r_0)_n + \sum_{m=1}^{\infty}\Bigl[(a_m)_n \cos( \frac{2 m \pi}{\beta} \tau ) + (b_m)_n \sin(\frac{2 m \pi}{\beta} \tau ) \Bigr] \,$   \cite{roy84} 
whereby a set of coefficients corresponds to a specific configuration for the n-th base pair and provides a measure of the fluctuational distance between the pair mates. The expansion also defines the integration measure $\oint {D}r_n$  over the space of the Fourier coefficients \cite{io03,io05} :

\begin{eqnarray}
& &\oint {D}r_{n} \equiv  {\frac{1}{\sqrt{2}\lambda_{cl}}} \int_{-\Lambda_{n}^{0}(T)}^{\Lambda_{n}^{0}(T)} d(r_0)_n \prod_{m=1}^{\infty}\Bigl( \frac{m \pi}{\lambda_{cl}} \Bigr)^2 \int_{-\Lambda_{n}(T)}^{\Lambda_{n}(T)} d(a_m)_n \int_{-\Lambda_{n}(T)}^{\Lambda_{n}(T)} d(b_m)_n \, \, , \, \nonumber
\\
\label{eq:31}
\end{eqnarray}

where $\lambda_{cl}$ is the classical thermal wavelength \cite{io14c},  $\Lambda_{n}^{0}(T)$ and $\Lambda_{n}(T)$ are the temperature dependent cutoffs for the radial fluctuations of the $n-th$ base pair. While the latter can be generally estimated by the formal condition, associated to the finite temperature path integral method, that the measure in Eq.~(\ref{eq:31}) normalizes the free particle action \cite{io11a}, I use hereafter the values obtained via a physically grounded argument recently devised to determine the cutoffs for a molecule in a specific twist conformation with a given set of model potential parameters \cite{io21}. 

{Note that the Fourier series and the associated integration measure have been taken only for the radial coordinate whereas there is no path expansion for the angular variables. Accordingly, the latter do not depend on $\tau$ and are integrated in a conventional way in Eq.~(\ref{eq:32}). This avoids to introduce arbitrary temperature dependent cutoffs for the angular variables and, a posteriori, explains why the kinetic energy in Eq.~(\ref{eq:01}) contains only radial contributions. 
}

\section*{5. Twist-stretch }

To address the twist-stretch relations,  the average helical repeat is computed by performing integrations over the ensemble of base pair configurations defined by  Eq.~(\ref{eq:32}). The idea behind the computational method is the following: the $n-$th twist fluctuation $\theta_{n}$ is measured with respect to the ensemble averaged twist, \, $<\theta_{n - 1}>$, obtained for the preceding base pair in the chain and augmented by $2\pi / h$ to account for the right-handedness of both the A- and B-forms.
$h$ is assumed to vary within a physically meaningful range, say \, $h_j  \in \, [h_{min}, \, h_{max}],   \, (j=1,...,J)$ \, and, for each $h_j$ in the range, the ensemble averaged twist angles $< \theta_n >$ are recursively computed by 
integrating over a twist fluctuation $\theta_{n}^{fl}$ around the value $\, <\theta_{n - 1}>  + 2\pi / h_j \,$.  

Setting,  $< \theta_1 >=\,0$ for the first base pair in the chain, the explicit formula for $< \theta_n >$ reads:

\begin{eqnarray}
& &< \theta_n >_{(n \geq 2)} =\,  < \theta_{n-1} > + \frac{2\pi}{h_j}  + \frac{\int_{-\theta_{M}}^{\theta_{M}} d \theta_{n}^{fl} \cdot ( \theta_n^{fl} ) \int_{- \phi _{M}}^{\phi _{M}} d \phi_n  \oint Dr_{n} \exp \bigl[- A_b [r_n, r_{n-1}, \phi_n, \theta_n]  \bigr]}{ \int_{-\theta_{M}}^{ \theta_{M}} d \theta_{n}^{fl} \int_{- \phi _{M} }^{\phi _{M} } d \phi_n  \oint Dr_{n} \exp \bigl[- A_b [r_n, r_{n-1}, \phi_n, \theta_n]  \bigr]} \, , \nonumber
\\
\label{eq:09}
\end{eqnarray}

and, from Eq.~(\ref{eq:09}), the average helical repeat is obtained as:

\begin{eqnarray}
< h >_j=\,\frac{2\pi N}{< \theta_N >} \, .
\label{eq:10}
\end{eqnarray}

Since the calculation is repeated for any input value $h_j$, the program generates a set of ensemble averaged values $\{ < h >_{j} \,  \} $ which 
define different twist conformations for the molecule subjected to a load. For any $< h >_{j}$, the corresponding free energy is calculated from Eq.~(\ref{eq:32}) as, $F=\, -\beta ^{-1} \ln Z_N$. By minimizing $F$ over the set of $J$ values, one finally selects the equilibrium twist conformation, denoted hereafter by $< h >_{j^{*}}$, for a given load. By tuning $F_{ex}$,   the twist profiles are derived for a specific helical molecule. 

Furthermore, for any twist  $< h >_j$, one calculates the average dimer distance $< d >_j$ by summing over the ensemble averaged  intra-strand base pairs distances i.e.,

\begin{eqnarray}
& &< d >_j= \, \frac{1}{N-1} \sum _{n=2}^{N} < \overline{d_{n,n-1}} > \, , \nonumber
\\
& &< \overline{d_{n,n-1}} > =\,  \frac{\oint Dr_{n}  \int_{-\theta_{M}}^{\theta_{M}} d \theta_{n}^{fl}   \int_{ -\phi _{M}}^{\phi _{M}} d \phi_n  \cdot \overline{d_{n,n-1}} \exp \bigl[- A_b [r_n, r_{n-1}, \phi_n, \theta_n]  \bigr]}{ \int_{-\theta_{M}}^{\theta_{M}} d \theta_{n}^{fl} \int_{-\phi _{M} }^{\phi _{M} } d \phi_n  \oint Dr_{n}  \exp \bigl[- A_b [r_n, r_{n-1}, \phi_n, \theta_n]  \bigr]} \, , \nonumber
\\
\label{eq:11}
\end{eqnarray}

whereby the equilibrium average distance $< d >_{j*}$ corresponds to the equilibrium $< h >_{j^{*}}$ conformation.  This permits to evaluate the size of the stretching induced by the external load both for the A- and B- type helical molecules.

While, for kilo-base B-DNA in solution under physiological condition, the experimental average helical repeat is estimated as $h^{exp} \sim 10.5$ \cite{wang79}, short DNA chains may have twist conformations which differ significantly from those of long chains. To account for these effects I first sample a broad range of $J \sim 70$ twist conformations around $h^{exp}$ in the absence of applied forces and then repeat the sampling procedure for any $F_{ex}$. {The forces are in the pico-Newton regime hence, they can oppose the bending and kinking caused by the buffeting of the solvent bath and ultimately straighten the helix. In fact, at the nano-scale, the room temperature thermal energy per nano-meter is $k_BT / nm \sim 4 pN$.}

\section*{6. Results and Discussion}

The theory is tested on a short  chain of $N = 21$ base pairs  which suffices to allow for about two turns of the helix. Despite the stacking potential in Eq.~(\ref{eq:02}) contains only two particles interactions, the helical conformation leads to base pairs correlations along the molecule backbone. I take the chain as homogeneous i.e., $D \equiv D_n$, $b \equiv b_n$, $K \equiv \,K_{n, n-1}$, $\rho  \equiv \,\rho _{n, n-1}$, $\alpha  \equiv \,\alpha _{n, n-1}$ and set the potential parameters as usually done in mesoscopic Hamiltonian investigations of the DNA properties \cite{io20,campa98}. Moreover, as short chains are subjected to sizeable end fraying effects \cite{zgarb14}, I assume that the stacking parameters for the end dimers containing the terminal base pairs, are one half of the value taken for the internal dimers. 

The same parameter values are used to model both the A- and B-type chains although earlier parametrization studies carried out with a one-dimensional mesoscopic model suggest that the RNA force constants may differ from those appropriate to DNA \cite{weber13}. However, as this study intends to highlight the effects brought about by the type of helical structure on the flexibility properties, I deem appropriate to ignore those details which may be ascribed to different choices of model potential parameters.

\begin{figure}
\includegraphics[height=11.0cm,width=9.0cm,angle=-90]{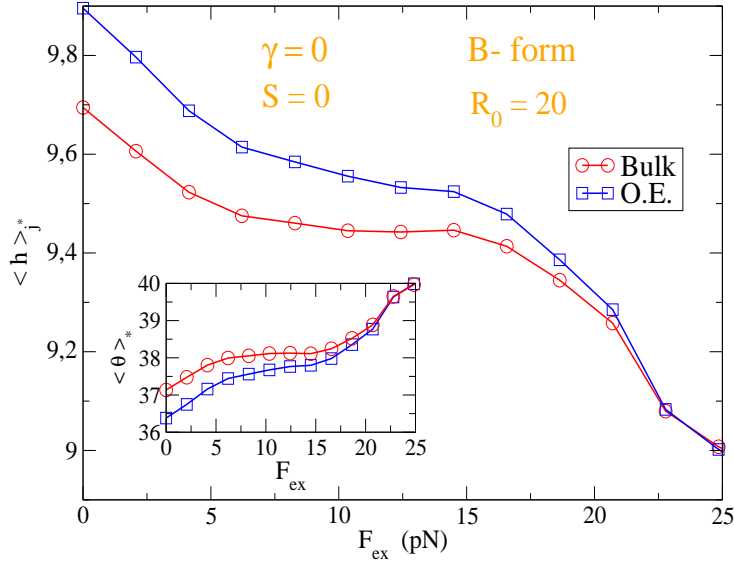}
\caption{\label{fig:3}(Color online)  Ensemble averaged helical repeat versus external load for the B- form helix. The molecule is considered both with (Open Ends) and without (bulk) terminal base pairs.  The inset shows the average twist angle for both cases. The base pair planes are taken perpendicular to the helical axis.
}
\end{figure}

I first focus on the standard B-form helix with neither tilt nor slide and an average diameter $R_0=\,20 \,\AA$.  To emphasize the role of terminal base pairs, Fig.~\ref{fig:3} displays the ensemble averaged equilibrium helical repeat versus the applied load in two cases: a) the full open ends (O.E.) chain made of $N-1$ dimers and b) the bulk of the chain made of $N-3$ dimers. In the former case, $< h >_{j^{*}}$ is significantly larger signaling that the terminal base pairs strongly affect the overall helix untwisting thus yielding an enhanced flexibility. This holds both in the absence of loads and for moderate loads up to about $20 \,pN$ whereas, for strong $F_{ex}$, the chain end effects tend to vanish as the over-twisting is more pronounced. The inset shows the equilibrium twist angles calculated for each $< h >_{j^{*}}$ both for the a) and the b) case.

\begin{figure}
\includegraphics[height=11.0cm,width=9.0cm,angle=-90]{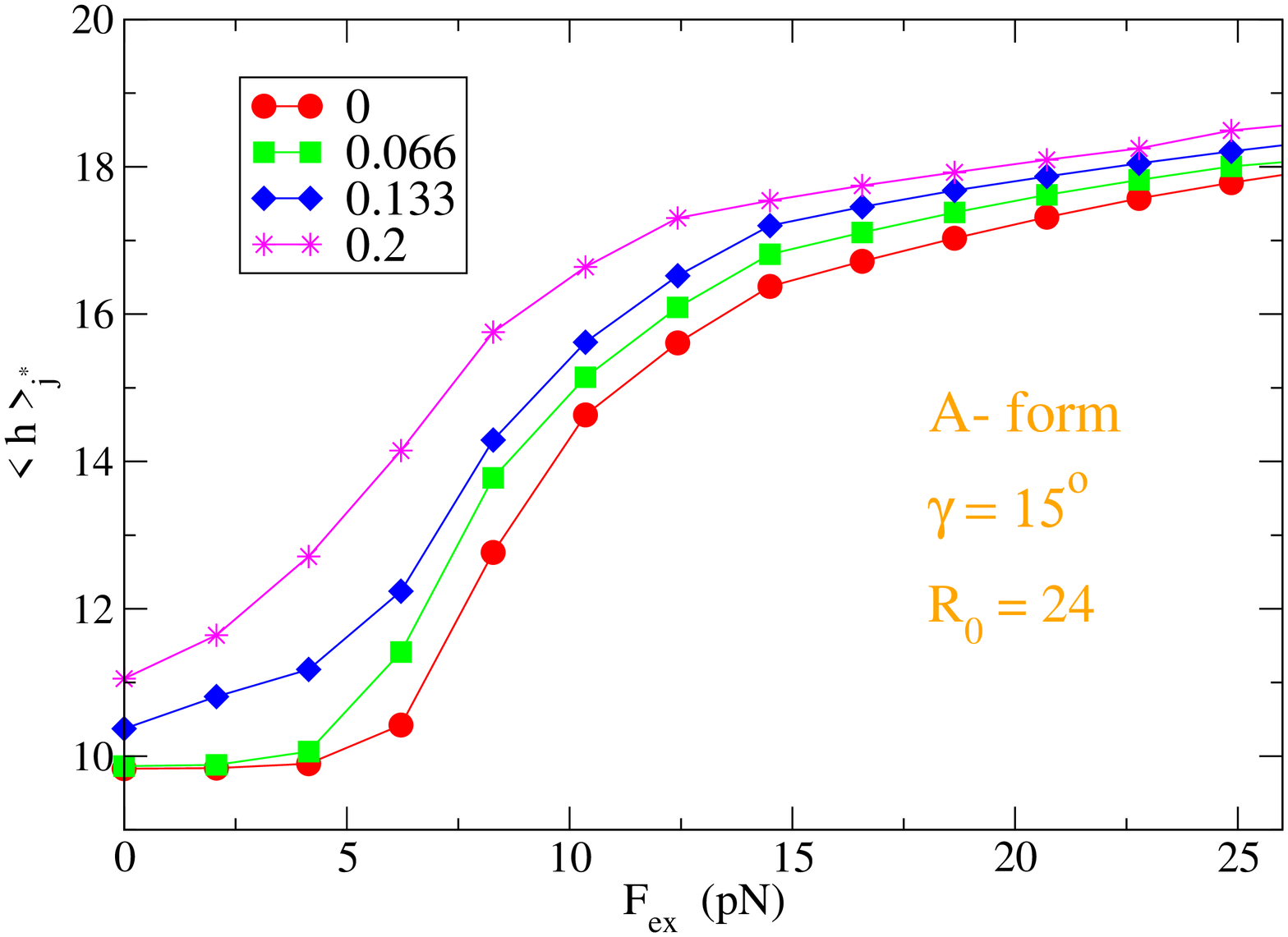}
\caption{\label{fig:4}(Color online)  Ensemble averaged helical repeat versus external load for the A- form helix. $\gamma $ is the inclination of the base pair planes with respect to the helical axis. Both the zero slide configuration and three configurations with finite $|S| / d$ values are considered. See Fig.~\ref{fig:2}.
}
\end{figure}

The general picture changes drastically once the A-form helix (with $R_0=\,24 \,\AA$) is considered as shown in Fig.~\ref{fig:4}. For the open ends chain, the ensemble averaged  $< h >_{j^{*}}$ is plotted versus $F_{ex}$ assuming a base pair inclination $\gamma =\, 15^{o}$ with respect to the helical axis consistent with X-ray diffraction data and molecular dynamics simulations \cite{dick83,zachar15}. Both the conformation with zero slide and three conformations with finite slide values are displayed. All plots show that $< h >_{j^{*}}$ grows under the effect of the stretching load revealing that the base pair inclination is the primary cause of the helix untwisting. The effect of the slide is superimposed to that of the inclination: by increasing $|S|$, the helix untwisting is larger for all $F_{ex}$ suggesting that the structural features of the dimers determine the overall flexibility of the chain. Similarly to Fig.~\ref{fig:3}, also for the A-form helix, the bulk helical repeat (not displayed) is generally smaller than the open end helical repeat calculated for the same parameters set. Likewise, the equilibrium twist angles can be straightforwardly calculated for each $< h >_{j^{*}}$.

\begin{figure}
\includegraphics[height=11.0cm,width=9.0cm,angle=-90]{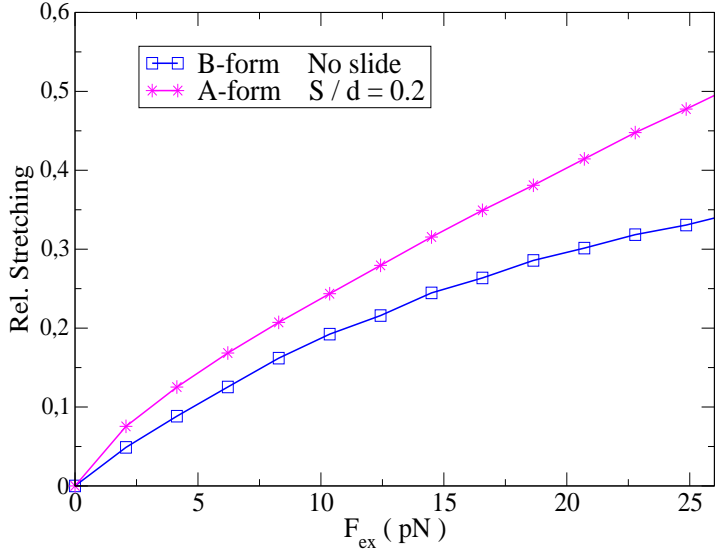}
\caption{\label{fig:5}(Color online)  Relative average elongation of the intra-dimer distance as a function of the load both for the B- form and A-form helices. The calculated relative stretching is precisely: \, $\bigl[< d >_{j*}(F_{ex}) / < d >_{j*}(F_{ex}=0)\bigr] - 1$. For the A-form, the inclination angle is $\gamma=\,15^{o} $. 
}
\end{figure}

In fact, the opposite twist-stretch patterns obtained in Fig.~\ref{fig:3} and Fig.~\ref{fig:4} should generate distinct stretching properties for the two helical forms.
This is precisely seen in Fig.~\ref{fig:5} wherein the average stretching between the base pairs in a dimer is calculated versus the applied load both for the B- form and A-form helices. In the latter case, I take the largest among the slide values in Fig.~\ref{fig:4}. While the average dimer distance is enhanced in both helices by increasing the load, the A-form helix stretches more than the B-form at any force and the relative stretching is found to grow almost linearly with $F_{ex}$.
This result is in line with optical tweezers measurements yielding a ds-RNA stretch modulus almost twofold lower than that of ds-DNA, albeit for kilo-base long molecules \cite{gonz13}.

\begin{figure}
\includegraphics[height=11.0cm,width=9.0cm,angle=-90]{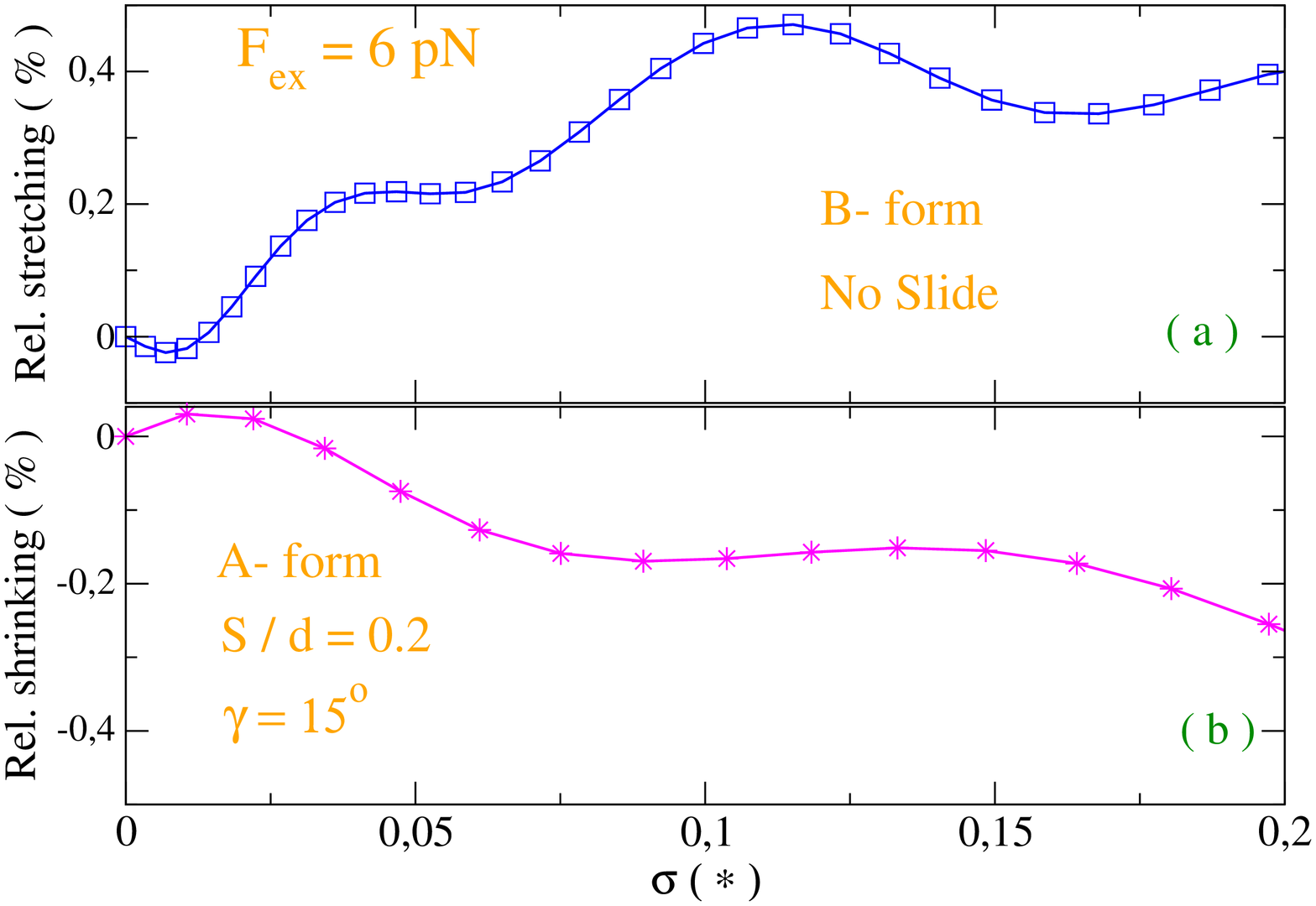}
\caption{\label{fig:6}(Color online) Assuming a fixed load $F_{ex}$, the helix over-twisting with respect to the equilibrium configuration is simulated for (a) the B- form and (b) A-form helix. The percentage relative stretching (a) and relative shrinking (b) of the intra-dimer distance is plotted versus the superhelical density 
$\sigma(*)$. 
}
\end{figure}

Finally, I simulate an experimental setup in which the helix is over-twisted under a constant load, e.g. $F_{ex}=\, 6 \,pN$, as shown in Fig.~\ref{fig:6}. This is done in magnetic tweezers assays \cite{croq06} in which a tethered molecule is stretched at constant force while rotating magnets add a torsional strain thereby changing the twist number with respect to the equilibrium value i.e., $(Tw)_{j*}=\, N /< h >_{j^{*}}$. This generates a superhelical density 
$\sigma(*)= (\Delta Tw)_{j*} / (Tw)_{j*}$ whereby  $(\Delta Tw)_{j*}$ measures the applied torsional strain i.e., the number of turns added to the helical twist, at fixed force. Under this conditions, the average intra-dimer distance is calculated by varying $\sigma(*)$ for the two helical forms. It is found that the behavior is opposite: while the extension of the B-form helix increases, the A-form shrinks when over-twisted.
The calculated stretching and shrinking in Fig.~\ref{fig:6} are not monotonous functions as, for any added turn, {the characteristic functions $< h >$ of the A- and B-forms differ from the respective $< h >_{j^{*}}$ hence, the intra-dimer distances do not correspond to a free energy minimum in our computational method.} This also causes the small oscillations observed at $\sigma(*)$ slightly larger than zero. Nevertheless the general trend, stretching of the B-form and shrinking of the A-form, is evident once the helices are markedly over-twisted hence, their average helical repeat values $< h >_{j}$ become significantly smaller than the equilibrium value $< h >_{j^{*}}$.
Taking the  specific $(Tw)_{j*}$ for the cases in Figs.~\ref{fig:6}(a),(b), I obtain that: (a) the over-winding of the B-form lenghtens the helix by an average slope of $1.67 \, nm$ per helix turn and (b) the over-winding of the A-form shortens the helix by an average $-1.61 \, nm$ per turn. Although these figures are larger (in absolute value, by about a factor two and four respectively) than those reported for ds-DNA and ds-RNA  \cite{dekker14} with a similar range of stretching forces (around $6 pN$), it should be noticed that our fragments are much shorter ($N=\,21$) than the kilo base long sequences taken in single-molecule magnetic tweezers experiments. While it is plausible that such short fragments may be stretched more than kilo base long molecules, quantitative comparison with available experimental data seems premature at this stage.

\section*{7. Conclusions}

The main conclusion of this study is that the opposite twist-stretch patterns  of ds-RNA and standard ds-DNA,  together with their stretching properties under constant tension, can be ascribed to specific structural differences in their di-nucleotide steps, piled along the helical axis. Incorporating these structural features in a 3D mesoscopic Hamiltonian model with angular degrees of freedom, I have used a  method based on path integral techniques to compute the average helical repeat,  both for the A- and B- form, of short fragments. Mesoscopic approaches are advantageous at such short length scales, whereas it has been pointed out that the elastic properties of nucleic acids are poorly described by the  worm-like-chain models of polymer physics which are instead traditionally applied to kilo-base long molecules. The path integral method permits to derive significant twist-stretch relations as a function of the structural parameters after selecting the equilibrium helical conformation by free energy minimization. Moreover, keeping the external load at a fixed value, I have applied a torsional strain to the molecule, studied the stretching response in a range of over-twisted configurations and performed quantitative estimates both for the lengthening of B-DNA and the shortening of RNA homogeneous short fragments.
The computational method can be further applied to test the elastic response of heterogeneous (and longer) sequences to external loads.


\vskip 1.5cm
















\begin{thebibliography}{widest-label}

\bibitem{stasiak97}
Kiianitsa K,  Stasiak A, (1997) Helical repeat of DNA in the region of homologous pairing. \emph{Proc Natl Acad Sci USA}  \textbf{94}: 7837-7840 

\bibitem{kalos11}
Apostolaki A,     Kalosakas G, (2011)  Targets of DNA-binding proteins in bacterial promoter regions present
enhanced probabilities for spontaneous thermal openings.  \textit{Phys  Biol }   \textbf{8}: 026006   

\bibitem{ort14}
Hudson W H,  Ortlund E A, (2014) The structure, function and evolution of proteins that bind DNA and RNA.  \textit{Nat Rev Mol Cell Biol}   \textbf{15}: 749-760 

\bibitem{marko15}
Marko J F,  (2015) Biophysics of protein–DNA interactions and chromosome organization. \textit{Physica A}  \textbf{418}: 126-153 

\bibitem{biton18}
 Biton Y Y, (2018) Eﬀects of Protein-Induced Local Bending and Sequence Dependence on the Conﬁgurations of Supercoiled DNA Minicircles. \textit{J Chem Theory Comput}    \textbf{14}:   2063-2075 

\bibitem{grzy19}
 Balcerak A,  Trebinska-Stryjewska A,  Konopinski R, Wakula M, Grzybowska E A, (2019) RNA-protein interactions: disorder, moonlighting and junk contribute to eukaryotic complexity.  \textit{Open Biol} \textbf{9}: 190096 

\bibitem{lee19}
 Choi S R,  Kim N H,  Jin H S,  Seo Y J,  Lee J,  Lee J H, (2019) Base-pair Opening Dynamics of Nucleic Acids in Relation to Their Biological Function. \textit{Comput Struct Biotech J} \textbf{17}:  797-804 

\bibitem{busta92}
 Smith S,  Finzi L,  Bustamante C,  (1992) Direct mechanical measurement of the elasticity of single DNA molecules by using magnetic beads. \emph{Science}   \textbf{258}:  1122-1126 


\bibitem{block97} 
 Wang M D,  Yin H,  Landick R,  Gelles J,  Block S M, (1997)  Stretching DNA with Optical Tweezers. \textit{Biophys J}    \textbf{72}: 1335-1346  

\bibitem{mameren09} 
 van Mameren J,  Gross P,  Farge G,  Hooijman P,  Modesti M,  Falkenberg M, Wuite G J L,  Peterman E J G,   (2009)  Unraveling the structure of DNA during overstretching by using multicolor, single-molecule fluorescence imaging.  \emph{Proc Natl Acad Sci USA}  \textbf{106}:  18231-18236 


\bibitem{cluzel96}
 Cluzel P,  Lebrun A,  Heller C,  Lavery R,  Viovy J L,  Chatenay D,  Caron F,  (1996) DNA: an extensible molecule. \emph{Science}   \textbf{271}:  792-794 

\bibitem{marko99}
 L\'{e}ger J F,   Romano G,   Sarkar A,   Robert J,   Bourdieu L,   Chatenay D,   Marko J F, (1999) Structural Transitions of a Twisted and Stretched DNA Molecule. \emph{Phys Rev Lett}  \textbf{83}: 1066 


\bibitem{rouz01}
 Rouzina I,  Bloomfield V A,  (2001) Force-Induced Melting of the DNA Double Helix 1. Thermodynamic Analysis. \emph{Biophys J}   \textbf{80}: 882-893 


\bibitem{maiti09}
 Santosh M,  Maiti P K, (2009) Force induced DNA melting. \textit{J Phys: Condens Matter} \textbf{21}: 034113 

\bibitem{nord14}
 Bosaeus N,  El-Sagheer A H,  Brown T, \r{A}kerman  B,  Nord\`{e}n  B, (2014) Force-induced melting of DNA—evidence for peeling and internal melting from force spectra on short synthetic duplex sequences. \textit{Nucleic Acids Res}  \textbf{42}:  8083-8091 

\bibitem{lomb14}
 Bongini L,  Melli L,  Lombardi V,  Bianco P, (2014) Transient kinetics measured with force steps discriminate between double-stranded DNA elongation and melting and define the reaction energetics. \textit{Nucleic Acids Res}  \textbf{42}:  3436-3449 


\bibitem{busta06} 
Gore J,  Bryant Z,  N\"{o}llmann M,  Le M U,  Cozzarelli N R, Bustamante  C,  (2006) DNA overwinds when stretched. \textit{ Nature}  \textbf{442}: 836-839 



\bibitem{tan15} 
 Wu Y Y,  Bao L,  Zhang X,  Tan Z J,  (2015)  Flexibility of short DNA helices with finite-length effect: from base pairs to tens of base pairs. \emph{J Chem Phys}  \textbf{142}: 125103 



\bibitem{dekker14}
 Lipfert J,  Skinner G M,  Keegstra J M,  Hensgens T,  Jager T,  Dulin D, K\"{o}ber M,  Yu Z,  Donkers S P,  Chou F C,  Das R,  Dekker  N H, (2014) Double-stranded RNA under force and torque: Similarities to and striking differences from double-stranded DNA. \textit{Proc Natl Acad Sci USA} {\bf 111}: 15408-15413 

\bibitem{frank53}
 Franklin R E,   Gosling R G,  (1953)  Molecular Configuration in Sodium Thymonucleate. \textit{Nature} \textbf{171}: 740-741 



\bibitem{puddu15}
 Grass R N,  Heckel R,  Puddu M,  Paunescu D,  Stark W J, (2015) Robust chemical preservation of digital information on DNA in silica with error-correcting codes. \textit{Angew Chem Int Ed}  \textbf{54}: 2552 


\bibitem{sena18}
 Ghoshdastidar D,  Senapati S, (2018) Dehydrated DNA in B-form: ionic liquids in rescue. \textit{Nucleic Acids Res} \textbf{46}: 4344-4353  

\bibitem{olson99}
 Kosikov K M,  Gorin A A,   Zhurkin V B,  Olson W K, (1999) DNA Stretching and Compression: Large-scale Simulations of Double Helical Structures. \textit{J Mol Biol} \textbf{289}: 1301-1326 

\bibitem{oroz10}
 Faustino I,  P\'{e}rez A,   Orozco M, (2010) Toward a Consensus View of Duplex RNA Flexibility. \textit{ Biophys J} \textbf{99}: 1876-1885 


\bibitem{tan16}
 Bao L,  Zhang X,  Jin L,   Tan Z J, (2016) Flexibility of nucleic acids: From DNA to RNA. \textit{Chin Phys B}  \textbf{25}:   018703 

\bibitem{bohr11}
 Olsen K,  Bohr J, (2011) The geometrical origin of the strain-twist coupling in double helices. \textit{AIP Advances} \textbf{1}: 012108 

\bibitem{gole} 
 Noy A,  Golestanian R, (2012) Length Scale Dependence of DNA Mechanical Properties. \textit{Phys Rev Lett}   {\bf 109}: 228101  


\bibitem{widom}
 Cloutier T E,  Widom J,  (2004)  Spontaneous Sharp Bending of Double-Stranded DNA. \emph{Mol Cell}  \textbf{ 14}: 355-362  

\bibitem{archer06}
 Yuan C,  Rhoades E,  Lou X W,   Archer L A, (2006) Spontaneous sharp bending of DNA: role of melting bubbles. \emph{Nucleic Acids Res}   \textbf{34}:  4554-4560  

\bibitem{wigg06} 
 Wiggins P A,  Heijden T V D,  Moreno-Herrero F,  Spakowitz A,  Phillips R,  Widom J,  Dekker C,  Nelson  P C, (2006) High flexibility of DNA on short length scales probed by atomic force microscopy. \textit{Nat Nanotechnol} \textbf{1}: 137 

\bibitem{vafa12}
 Vafabakhsh R,  Ha T, (2012) Extreme Bendability of DNA Less than 100 Base Pairs Long Revealed by Single-Molecule Cyclization. \emph{Science}  \textbf{337}:   1097-1101   


\bibitem{maiti15}
 Garai A,   Saurabh S,   Lansac Y,  Maiti P K, (2015) DNA Elasticity from Short DNA to Nucleosomal DNA. \textit{J Phys Chem B}  \textbf{119}: 11146-11156 


\bibitem{io16b}
 Zoli M,  (2016) J- factors of short DNA molecules. \textit{J Chem Phys}   {\bf 144}:   214104  

\bibitem{lam17}
Lam P M, Zhen Y, (2017) Cyclization of short DNA fragments. \textit{Physica A} {\bf 482}: 569



\bibitem{zachar15}
 Liebl K,  Drsata T,  Lanka\u{s} F,  Lipfert J,   Zacharias M, (2015)  Explaining the striking difference in twist-stretch
coupling between DNA and RNA: A comparative molecular dynamics analysis. \textit{Nucleic Acid Res}  \textbf{43}:  10143-10156  


\bibitem{io18b}
 Zoli M, (2018) Short DNA persistence length in a mesoscopic helical model. \textit{EPL - Europhysics Letters} {\bf 123}:   68003 

\bibitem{zhang23}
 Zhang Y,  He L,  Li S,  (2023) Temperature dependence of DNA elasticity: An all-atom molecular dynamics simulation study. \textit{J Chem Phys} {\bf 158}:  094902 


\bibitem{io19}
Zoli M, (2019) DNA size in confined environments. \textit{Phys Chem Chem Phys}   {\bf 21}:   12566  

\bibitem{io20b}
Zoli M,  (2020) Stretching DNA in hard-wall potential channels. \textit{EPL - Europhysics Letters} {\bf 130}: 28002 

\bibitem{pey89}
Peyrard M,  Bishop  A R, (1989) Statistical mechanics of a nonlinear model for DNA denaturation. \textit{Phys Rev Lett}    {\bf 62}: 2755

\bibitem{barbi99}
 Barbi M,  Cocco S,  Peyrard M, (1999) Helicoidal model for DNA opening. \textit{Phys Lett A} \textbf{253}:  358 

\bibitem{io22}
Zoli M, (2022) Non-linear Hamiltonian models for DNA.  \textit{European Biophys J}  \textbf{51}: 431-447 

\bibitem{io16a}
Zoli M,  (2016) Flexibility of short DNA helices under mechanical stretching. \textit{Phys Chem Chem Phys} {\bf 18}:  17666 

\bibitem{carlo06}
Fohrer J,  Hennig M,  Carlomagno T, (2006) Influence of the 2'-hydroxyl group conformation on the stability of A-form helices in RNA. \textit{J Mol Biol}  \textbf{356}: 280-287 

\bibitem{calla}
 Calladine C R,  Drew H R,  (1992) \emph{Understanding DNA}, (Academic Press, San Diego) 


\bibitem{dick83}
 Dickerson R E, (1983) The DNA Helix and How It Is Read. \textit{Scientific American}  \textbf{249}: 94-111 


\bibitem{romano13} 
 Romano F,   Chakraborty D,   Doye J P K,  Ouldridge T E,   Louis A A,   (2013) Coarse-grained simulations of DNA overstretching. \emph{J Chem Phys}  \textbf{138}: 085101 

\bibitem{pey93}
Dauxois T,  Peyrard M,  Bishop A R,  (1993) Entropy driven DNA denaturation. \emph{Phys. Rev. E}  \textbf{47}:   R44-R47  

\bibitem{pey09}
Peyrard M, Cuesta-L\'{o}pez S, James G, (2009) Nonlinear analysis of the dynamics of DNA breathing. \textit{J Biol Phys} \textbf{35}:  73


\bibitem{coll95}
Zhang F,  Collins M A, (1995) Model simulations of DNA dynamics. \textit{Phys Rev E} {\bf 52}: 4217 

\bibitem{druk01}
 Drukker K,  Wu G,  Schatz G C,  (2001)  Model simulations of DNA denaturation dynamics. \emph{J. Chem. Phys.} \textbf{114}:  579-590  

\bibitem{zdrav01}
 Zdravkovi\'{c} S,   Satari\'{c} M V, (2001) The Impact of Viscosity on the DNA Dynamics. \textit{Phys Scr} \textbf{64}: 612

\bibitem{wart85}
 Wartell R M,  Benight A S, (1985) Thermal denaturation of DNA molecules: a comparison of theory with experiment. \textit{Phys Rep} \textbf{126}:   67-107 

\bibitem{eijck11}
 van Eijck L,  Merzel F,  Rols S,  Ollivier J,  Forsyth VT,  Johnson MR, (2011) Direct Determination of the Base-Pair Force Constant of DNA from the
Acoustic Phonon Dispersion of the Double Helix.  \textit{Phys Rev Lett}  \textbf{ 107}:  088102 



\bibitem{kalos20}
 Hillebrand M,  Kalosakas G,   Skokos Ch,   Bishop A R, (2020) Distributions of bubble lifetimes and bubble lengths in DNA. \emph{Phys Rev E} \textbf{102}: 062114 

\bibitem{io20}
Zoli M, (2020) First-passage probability: a test for DNA Hamiltonian parameters. \textit{Phys Chem Chem Phys} {\bf 22}:  26901 

\bibitem{io18c}
Zoli M,  (2018)  End-to-end distance and contour length distribution functions of DNA helices.   \emph{J Chem Phys}  \textbf{148}:   214902 

\bibitem{fehi}
 Feynman R P,   Hibbs A R, (1965)  {\it Quantum Mechanics and Path Integrals}, (Mc Graw-Hill, New York)  

\bibitem{io14b}
Zoli M,  (2014) Entropic Penalties in Circular DNA Assembly. \textit{J Chem Phys}  {\bf 141}: 174112 

\bibitem{kleinert}
 Kleinert H, (2004) {\it Path Integrals in Quantum Mechanics, Statistics,
Polymer Physycs and Financial Markets}, ( World Scientific Publishing, Singapore )


\bibitem{zhang97}
 Zhang Y L,  Zheng W M,  Liu J X,   Chen Y Z,  (1997) Theory of DNA melting based on the Peyrard-Bishop model. \textit{ Phys Rev E}  \textbf{56}:  7100-7115  

\bibitem{hando12}
 Sulaiman A,  Zen F P,  Alatas H,  Handoko  L T, (2012) The thermal denaturation of the Peyrard-Bishop model with an external potential. \textit{Phys Scr} \textbf{86}:  015802 

\bibitem{albu14}
 Macedo D X,  Guedes I,  Albuquerque E L,  (2014)  Thermal properties of a DNA denaturation with solvent interaction. \textit{Physica A}  \textbf{404}: 234-241 

\bibitem{singh15}
 Singh A, Singh N,  (2015) Effect of salt concentration on the stability of heterogeneous DNA. \emph{Physica A} \textbf{419}: 328-334  

\bibitem{lank10}
 Lanka\u{s} F,  \u{S}pa\u{c}kov\'{a} N,  Moakher M,  Enkhbayar P, \u{S}poner J, (2010) A measure of bending in nucleic acids structures applied to A-tract DNA. \emph{Nucleic Acids Res}   \textbf{38}: 3414-3422   


\bibitem{kim14}
 Le T T,  Kim H D,  (2014)  Probing the elastic limit of DNA bending. \emph{Nucleic Acids Res}   \textbf{42}:  10786-10794  

\bibitem{roy84}
Royer A, (1984) On the Fourier series representations of path integrals. \textit{J Math Phys} \textbf{25}: 2873 


\bibitem{io03}
Zoli M (2003)  Path Integral Description of a Semiclassical Su-Schrieffer-Heeger Model. \emph{Phys. Rev. B}  \textbf{67}:  195102


\bibitem{io05}
Zoli M (2005)  Path Integral of the Two Dimensional Su-Schrieffer-Heeger Model. \emph{Phys. Rev. B}  \textbf{71}:  205111


\bibitem{io14c}
Zoli M,  (2014) Twist versus Nonlinear Stacking in Short DNA Molecules.  \textit{J Theor Biol} {\bf 354}:  95-104 

\bibitem{io11a}
Zoli M,  (2011) Stacking Interactions in Denaturation of DNA Fragments.  \emph{Eur Phys J E} {\bf 34}: 68 


\bibitem{io21}
Zoli M,   (2021) Base pair fluctuations in helical models for nucleic acids.  \emph{J Chem Phys} {\bf 154}: 194102  


\bibitem{wang79}
 Wang J C, (1979) Helical repeat of DNA in solution. \emph{Proc Natl Acad Sci USA}   \textbf{76}:  200-203 


\bibitem{campa98}
 Campa A,  Giansanti A, (1998) Experimental tests of the Peyrard-Bishop model applied to the melting of very short DNA chains.  \textit{Phys Rev E}  \textbf{58}: 3585-3588  


\bibitem{zgarb14}
 Zgarbov\'{a} M,  Otyepka M,  Sponer J,  Lanka\u{s} F,   Jurecka P, (2014) Base Pair Fraying in Molecular Dynamics Simulations of DNA and
RNA. \textit{J Chem Theory Comput} \textbf{10}: 3177-3189 

\bibitem{weber13}
 Weber G,  (2013) Mesoscopic model parametrization of hydrogen bonds and stacking interactions of RNA from melting temperatures.  \emph{Nucleic Acids Res}  \textbf{41}: e30 

\bibitem{gonz13}
 Herrero-Gal\`{a}n E,  Fuentes-Perez M E,   Carrasco C,   Valpuesta J M,  Carrascosa J L,   Moreno-Herrero F,   Arias-Gonzalez J R, (2013) Mechanical Identities of RNA and DNA Double Helices Unveiled at the Single-Molecule Level.  \textit{J Am Chem Soc } \textbf{135}: 122-131 

\bibitem{croq06}
 Lionnet T,   Joubaud S,   Lavery R,   Bensimon D,  Croquette V,  (2006) Wringing out DNA. \emph{Phys Rev Lett}  \textbf{96}: 178102 










 
 
 


 











\end{thebibliography}
\end{document}